\documentclass[prb,twocolumn,showpacs]{revtex4}
\usepackage{graphicx}
\usepackage{dcolumn}
\usepackage{bm}
\usepackage{color}
\usepackage{amsmath}
\usepackage{amsfonts}
\usepackage{amssymb}

\input{tcilatex}

\begin{document}

\title{Imaginary eigenvalues of Hermitian Hamiltonian with an inverted
potential well and transition to the real spectrum at exceptional point by a
non-Hermitian interaction}
\author{Ni Liu}
\email{liuni2011520@sxu.edu.cn}
\affiliation{Institute of Theoretical Physics, State Key Laboratory of Quantum Optics and
Quantum Optics Devices, Shanxi University, Taiyuan 030006, Shanxi, China}
\author{Meng Luo}
\affiliation{Institute of Theoretical Physics, State Key Laboratory of Quantum Optics and
Quantum Optics Devices, Shanxi University, Taiyuan 030006, Shanxi, China}
\author{Zuohong Wang}
\affiliation{Institute of Theoretical Physics, State Key Laboratory of Quantum Optics and
Quantum Optics Devices, Shanxi University, Taiyuan 030006, Shanxi, China}
\author{J. -Q. Liang}
\email{jqliang@sxu.edu.cn}
\affiliation{Institute of Theoretical Physics, State Key Laboratory of Quantum Optics and
Quantum Optics Devices, Shanxi University, Taiyuan 030006, Shanxi, China}

\begin{abstract}
We in this paper study the hermiticity of Hamiltonian and energy spectrum
for the $SU(1,1)$ systems.\textbf{\ }The Hermitian Hamiltonian can possess
imaginary eigenvalues in\textbf{\ }contrast with the common belief that
hermiticity is a sufficient condition for real spectrum. The imaginary
eigenvalues are derived in algebraic method with imaginary-frequency boson
operators for the Hamiltonian of inverted potential well. Dual sets of
mutually orthogonal eigenstates are required corresponding respectively to
the complex conjugate eigenvalues. Arbitrary order eigenfunctions seen to be
the polynomials of imaginary frequency are generated from the normalized
ground-state wave functions, which are spatially non-localized. The
Hamiltonian including a non-Hermitian interaction term can be converted by
similarity transformation to the Hermitian one with an effective potential
of reduced slope, which is turnable by the interaction constant. The
transformation operator should not be unitary but Hermitian different from
the unitary transformation in ordinary quantum mechanics. The effective
potential vanishes at a critical value of coupling strength called the
exceptional point, where all eigenstates are degenerate with zero eigenvalue
and transition from imaginary to real spectra appears. The $SU(1,1)$
generator $\widehat{S}_{z}$ with real eigenvalues determined by the
commutation relation of operators, however, is non-Hermitian in the
realization of imaginay-frequency boson operators. The classical counterpart
of the quantum Hamiltonian with non-Hermitian interaction is a complex
function of the canonical variables. It becomes by the canonical
transformation of variables a real function indicating exactly the one to
one quantum-classical correspondence of Hamiltonians.
\end{abstract}

\maketitle
\date{\today }
\date{\today }
\date{\today }
\date{\today }
\date{\today }
\date{\today }
\date{\today }
\date{\today }

\section{Introduction}

In quantum mechanics the Hamiltonian has to be Hermitian to
guarantee real energy spectrum for the closed systems. The
non-Hermitian Hamiltonian describes usually open systems, which
exchange energy continuously with external source and thus possess
complex energy spectrum. Bender and his collaborators proposed for
the first time a creative finding that the Hermitian Hamiltonian is
sufficient but not necessary to have real eigenvalues
\cite{CM98,CM99,CM02}. According to them the non-Hermitian
Hamiltonian with real spectrum has the parity-time ($PT$) reflection
symmetry \cite{CM98,CM99,CM02,GuResult,GuAnn,LiuResult}.
Mostafazadeh named the non-Hermitian Hamiltonian with real spectrum
as pseudo-Hermitian, which satisfies a necessary condition
\cite{Mosta1,Mosta2,Mosta3}. While the $PT$ symmetry is neither
necessary nor sufficient condition for a non-Hermitian Hamiltonian
to have real spectrum. It is found that the $PT$ -asymmetric
Hamiltonian can also possess real spectra
\cite{GuResult,GuAnn,LiuPysica,Amaa}.

For the $PT$ -symmetric non-Hermitian Hamiltonians there exists an
exceptional point, which separates the unbroken $PT$ symmetry region with
pure real-spectrum and a broken $PT$ symmetry region, where eigenvalues are
complex \cite{CM98,CM04,CM07}. This transition at the exceptional point has
been observed in numerous laboratory experiments \cite{CM18,DN18} for
various non-Hermitian Hamiltonians. The non-Hermitian Hamiltonian has become
a rapidly developing subject in different branches of physics\cite%
{CH13,ZZ16,SB12,CM13,CE10,YW19} The spectrum property becomes complicated at
the exceptional points\cite{LC14,PG15,HX16,WL17} due to the symmetry
breaking, which can be modulated by dissipative media\cite%
{SK19,JZ18,YZ17,FQ18}. The exceptional point and complex domain of
eigenvalues have been well studied particularly in relation with optical
systems\cite{CM18,DN18}.

As a matter of fact the content of quantum mechanics is more rich than what
we are known. Not only the non-Hermitian Hamiltonians of open systems but
also the Hermitian Hamiltonian of a conservative system can have complex
spectrum if the potential is unbounded from below. The complex spectrum of
an inverted double-well potential was studied long ago in terms of the
instanton method under semiclassical approximation \cite{LM92,LM94}. The
imaginary part of energy eigenvalues characterizes the decay rate or life
time of metastable states \cite{LM92,LM94}. As of yet the rigorous complex
eigenvalues and associated eigenstates have not been given in the framework
of quantum mechanics for the Hermitian Hamiltonians.

The gain and loss of energy are not the only mechanism for complex
eigenvalues. We propose in the present paper a solvable Hamiltonian
consisting of $SU(1,1)$ generators, which can be realized by single-mode
boson operators. In the absence of external field the Hamiltonian, which is
Hermitian however with imaginary eigenvalues, describes a particle in an
inverted potential well. Classically the particle moves acceleratingly away
from the center of the inverted potential well. The quantum wave functions
are, of course, spatially non-localized. Inverted potential well unbounded
from below is found in the minisuperpace model of expanding universe, which
is established from Einstein equation by including the cosmology constant
\cite{Liang}.

We demonstrate in the present paper that the non-Hermitian interaction of%
\textbf{\ }an external field can decrease continuously the slope of inverted
potential well by the increase of coupling constant. At the end the
potential vanishes at a critical coupling-value, namely the exceptional
point, where all eigenstates are degenerate with a zero eigenvalue. Beyond
the exceptional point the eigenvalues become real as that of a normal
oscillator. We discover a peculiar transition from imaginary to real spectra
at the exceptional point induced by the non-Hermitian interaction.

The Hamiltonian with a non-Hermitian interaction term can be converted by a
similarity transformation to a Hermitian one. The transformation operator is
not unitary but Hermitian \cite{GuAnn} different from the unitary
transformation in the ordinary quantum mechanics. The classical counterpart
of the non-Hermitian Hamiltonian is seen to be a complex function of
canonical variables. It becomes under the canonical transformation of
variables a real function indicating strict one to one correspondence of the
quantum and classical Hamiltonians \cite{GuAnn}.

The article is organized as follows. In Sec.II, the imaginary spectrum and
eigenstates are obtained by means of the algebraic method for the Hermitian
Hamiltonian with an inverted potential well. We demonstrate in Sec.III the
transition from imaginary to real spectra at exceptional point by the
non-Hermitian interaction. The quantum-classical correspondence for the
Hermitian and non-Hermitian Hamiltonians is revealed in Sec.IV. The
conclusion and discussion are presented in Sec.V.

\section{Hermitian Hamiltonian with an inverted potential well and imaginary
spectrum}

The hermiticity of Hamiltonian is not sufficient condition for real
eigenvalues. We assume that the Hermitian Hamiltonian possesses discrete
complex eigenvalues. There must exist dual-set of eigenstates such that%
\begin{eqnarray}
\widehat{H}|u_{n}\rangle _{r} &=&E_{n}|u_{n}\rangle _{r},\quad _{r}\langle
u_{n}|\widehat{H}=_{r}\langle u_{n}|E_{n}^{\ast }  \notag \\
\widehat{H}|u_{n}\rangle _{l} &=&E_{n}^{\ast }|u_{n}\rangle _{l},\quad
_{l}\langle u_{n}|\widehat{H}=_{l}\langle u_{n}|E_{n},  \label{A1}
\end{eqnarray}%
in which the subscripts "$r$", "$l$" denote respectively the "ket" and "bra"
states. The two sets of eigenstates are mutually orthonormal
\begin{equation*}
_{l}\langle u_{n}|u_{m}\rangle _{r}=_{r}\langle u_{n}|u_{m}\rangle
_{l}=\delta _{n,m}.
\end{equation*}%
The density operators defined by $\widehat{\rho }=|\psi \rangle _{rl}\langle
\psi |$, $\widehat{\rho }^{\dag }=|\psi \rangle _{lr}\langle \psi |$ are
non-Hermitian invariants
\begin{equation*}
\frac{d\widehat{\rho }}{dt}=\frac{d\widehat{\rho }^{\dag }}{dt}=0.
\end{equation*}%
They obey the quantum Liouvill equation (in the unit convention $\hbar =1$
throughout the paper),%
\begin{equation*}
i\frac{\partial \widehat{\rho }}{\partial t}=\left[ \widehat{H},\quad
\widehat{\rho }\right] ,\quad i\frac{\partial \widehat{\rho }^{\dag }}{%
\partial t}=\left[ \widehat{H},\quad \widehat{\rho }^{\dag }\right] ,
\end{equation*}%
which are in consistence with the Schr\"{o}dinger equations%
\begin{equation*}
i\frac{\partial }{\partial t}|\psi \rangle _{\gamma }=\widehat{H}|\psi
\rangle _{\gamma },
\end{equation*}%
for both the "bra" and "ket" states ($\gamma =l,r$). The state densities $%
\langle x|\widehat{\rho }|x\rangle $, $\langle x|\widehat{\rho }^{\dag
}|x\rangle $ are conserved quantities satisfying the basic requirement of
quantum mechanics.

We consider the Hermitian Hamiltonian for a particle of unit mass in the
inverted potential well
\begin{equation}
\widehat{H}=\Omega \left( \frac{\widehat{p}^{2}}{2}-\frac{\widehat{x}^{2}}{2}%
\right) ,  \label{A}
\end{equation}%
in which $\widehat{x}$, $\widehat{p}$ are dimensionless operator with the
usual commutation relation $\left[ \widehat{x},\widehat{p}\right] =i$. The
classical orbit of the particle in the inverted potential well is simply%
\begin{equation*}
x\left( t\right) =\frac{v_{0}}{\Omega }\sinh (\Omega t),
\end{equation*}%
with initial position $x_{0}=0$ and velocity $v_{0}$. A minisuperspace model
\cite{Liang} of the universe is found also as an inverted potential well,
which results in the accelerating expansion. The Hamiltonian Eq.(\ref{A})
becomes with the imaginary frequency boson-operators
\begin{equation*}
\widehat{H}=i\Omega \left( \widehat{b}_{+}\widehat{b}_{-}+\frac{1}{2}\right)
,
\end{equation*}%
where%
\begin{equation}
\widehat{b}_{-}=\sqrt{\frac{i}{2}}\left( \widehat{x}+\widehat{p}\right) ,%
\widehat{b}_{+}=\sqrt{\frac{i}{2}}\left( \widehat{x}-\widehat{p}\right) .
\label{G}
\end{equation}%
The imaginary-frequency boson operators satisfy the same commutation
relation as the usual boson operators%
\begin{equation}
\left[ \widehat{b}_{-},\widehat{b}_{+}\right] =1.  \label{H}
\end{equation}%
The commutation relation of $SU(1,1)$ generators \cite{GuAnn,LLM} is
invariant under the realization of imaginary-frequency boson-operators
\begin{equation}
\widehat{S}_{z}=\frac{1}{2}\left( \widehat{b}_{+}\widehat{b}_{-}+\frac{1}{2}%
\right) ,\widehat{S}_{+}=\frac{1}{2}\left( \widehat{b}_{+}\right) ^{2},%
\widehat{S}_{-}=\frac{1}{2}\left( \widehat{b}_{-}\right) ^{2}  \label{C}
\end{equation}%
such that%
\begin{equation}
\left[ \widehat{S}_{z},\widehat{S}_{\pm }\right] =\pm \widehat{S}_{\pm },%
\left[ \widehat{S}_{+},\widehat{S}_{-}\right] =-2\widehat{S}_{z}.  \label{E}
\end{equation}%
Thus the Hamiltonian Eq.(\ref{A}), is represented as%
\begin{equation}
\widehat{H}=2i\Omega \widehat{S}_{z}.  \label{B}
\end{equation}%
Since%
\begin{equation*}
\widehat{S}_{z}^{\dag }=\frac{1}{2}\left( -\widehat{b}_{-}\widehat{b}_{+}+%
\frac{1}{2}\right) =-\widehat{S}_{z},
\end{equation*}%
the Hamiltonian Eq.(\ref{B}) is Hermitian as it should be. While the $%
SU(1,1) $ generator $\widehat{S}_{z}$ is no longer Hermitian different from
the representation of normal boson operators \cite{GuAnn,LLM}. The
imaginary-frequency boson-number operators
\begin{equation}
\widehat{n}_{I}=\widehat{b}_{+}\widehat{b}_{-};\quad \widehat{n}_{I}^{\dag
}=-\widehat{b}_{-}\widehat{b}_{+}  \label{B2}
\end{equation}%
are non-Hermitian with eigenstates given by%
\begin{eqnarray}
\widehat{n}_{I}|n\rangle _{r} &=&n|n\rangle _{r},_{r}\langle n|n=_{r}\langle
n|\widehat{n}_{I}^{\dag },  \label{B6} \\
\widehat{n}_{I}^{\dag }|n\rangle _{l} &=&n|n\rangle _{l},_{l}\langle
n|n=_{l}\langle n|\widehat{n}_{I}^{\dag }  \notag
\end{eqnarray}%
The orthogonal condition is%
\begin{equation}
_{l}\langle n|m\rangle _{r}=\delta _{nm}.  \label{B3}
\end{equation}%
The Hermitian Hamiltonian possesses discrete imaginary eigenvalues
\begin{equation}
\widehat{H}|n\rangle _{\gamma }=\pm i\Omega \left( n+\frac{1}{2}\right)
|n\rangle _{\gamma },  \label{B1}
\end{equation}%
with $\gamma =r,l$ respectively for the "ket" and "bra" states. Indeed the
Hermitian Hamiltonian with imaginary eigenvalues Eq.(\ref{B1}) has a
dual-set of eigenstates in agreement with the general theory Eq.(\ref{A1}).

Based on the commutation relation of imaginary-frequency boson-operators Eq.(%
\ref{H}) the arbitrary order eigenstates can be generated from the ground
states such that (see Appendix for the derivation)
\begin{eqnarray*}
|n\rangle _{r} &=&\frac{\left( \widehat{b}_{+}\right) ^{n}}{\sqrt{n!}}%
|0\rangle _{r}, \\
|n\rangle _{l} &=&\frac{\left( \widehat{b}_{-}\right) ^{n}}{(-i)^{n}\sqrt{n!}%
}|0\rangle _{l}.
\end{eqnarray*}%
From the ground-state equations%
\begin{equation*}
\widehat{b}_{-}|0\rangle _{r}=0;\widehat{b}_{+}|0\rangle _{l}=0
\end{equation*}%
the arbitrary order eigenfunctions are seen to be imaginary-frequency
polynomials
\begin{eqnarray}
&&\psi _{r,n}\left( x,\Omega \right)  \notag \\
&=&\frac{1}{\sqrt{n!}}\left( \sqrt{\frac{i}{2}}\right) ^{n}\left( \frac{i}{%
\pi }\right) ^{\frac{1}{4}}\left( x-i\frac{d}{dx}\right) ^{n}e^{-i\frac{x^{2}%
}{2}},  \notag \\
&&\psi _{l,n}\left( x,\Omega \right)  \notag \\
&=&\frac{1}{\left( -i\right) ^{n}\sqrt{n!}}\left( \sqrt{\frac{i}{2}}\right)
^{n}\left( \frac{-i}{\pi }\right) ^{\frac{1}{4}}\left( x+i\frac{d}{dx}%
\right) ^{n}e^{i\frac{x^{2}}{2}}  \label{C1}
\end{eqnarray}%
which derived analytically in the Appendix are spatially non-localized
different from the Hermit polynomials of the normal oscillator. It may be
better to remark that the wave functions are represented in the
dimensionless coordinate $x,$ which is scaled by the square root of
frequency $\sqrt{\Omega }.$ In the ordinary space coordinate, $x$ should be
replaced by $\sqrt{\Omega }x$ (see Appendix). Thus the eigenfunctions Eq.(%
\ref{C1}) are actually frequency dependent in the usual space coordinate.

\section{Transition from imaginary to real domains of spectrum and the
exceptional point}

We apply a non-Hermitian interaction to the Hamiltonian for a particle in
the inverted potential well

\begin{equation}
\widehat{H}=\widehat{H}_{0}+G\left( \widehat{S}_{+}-\widehat{S}_{-}\right) ,
\label{F}
\end{equation}%
where%
\begin{equation*}
\widehat{H}_{0}=2i\Omega \widehat{S}_{z}
\end{equation*}%
and $G$ is the coupling constant. $\widehat{H}_{0}$ is Hermitian however
with imaginary eigenvalue as shown in the previous section, while total
Hamiltonian is non-Hermitian,
\begin{equation}
\widehat{H}^{\dag }=\widehat{H}_{0}-G\left( \widehat{S}_{+}-\widehat{S}%
_{-}\right)  \label{F3}
\end{equation}%
where $\widehat{S}_{+}$, $\widehat{S}_{-}$ given in Eq.(\ref{C}) are
anti-Hermitian $\widehat{S}_{+}^{\dag }=-\widehat{S}_{+}$, $\widehat{S}%
_{-}^{\dag }=-\widehat{S}_{-}$ seen from the definition of boson operators
Eq.(\ref{G})$.$

The non-Hermitian Hamiltonian with complex eigenvalues can be also
transformed to a Hermitian one in general. To see this we assume that%
\begin{equation*}
\widehat{H}|u_{n}\rangle _{r}=\varepsilon _{n}|u_{n}\rangle _{r},\quad
\widehat{H}^{\dag }|u_{n}\rangle _{l}=\varepsilon _{n}^{\ast }|u_{n}\rangle
_{l}.
\end{equation*}%
Under a similarity transformation with a Hermitian operator $\widehat{R}%
^{\dag }=\widehat{R}$, we have
\begin{equation*}
\widehat{R}\widehat{H}\widehat{R}^{-1}|\widetilde{u}_{n}\rangle
_{r}=\varepsilon _{n}|\widetilde{u}_{n}\rangle _{r},\quad \widehat{R}^{-1}%
\widehat{H}^{\dag }\widehat{R}|\widetilde{u}_{n}\rangle _{l}=\varepsilon
_{n}^{\ast }|\widetilde{u}_{n}\rangle _{l},
\end{equation*}%
with%
\begin{equation}
|\widetilde{u}_{n}\rangle _{l}=\widehat{R}^{-1}|u_{n}\rangle _{l},\quad |%
\widetilde{u}_{n}\rangle _{r}=\widehat{R}|u_{n}\rangle _{r}.  \label{F1}
\end{equation}%
If the equality%
\begin{equation}
\widehat{R}\widehat{H}\widehat{R}^{-1}=\widehat{R}^{-1}\widehat{H}^{\dag }%
\widehat{R}\equiv \widetilde{H},  \label{F2}
\end{equation}%
is satisfied, the transformed Hamiltonian $\widetilde{H}=\widetilde{H}^{\dag
}$ is Hermitian with complex eigenvalues.

To this end the Hamiltonian Eq.(\ref{F}) can be diagonalized in terms of a
similarity transformation with the Hermitian operator
\begin{equation}
\widehat{R}=e^{-i\frac{\eta }{2}\left( \widehat{S}_{+}+\widehat{S}%
_{-}\right) },  \label{C3}
\end{equation}%
where the real-value parameter $\eta $ is to be determined. The $SU(1,1)$
generators are realized by the imaginay-frequency boson operators in Eq.(\ref%
{C}). It is easy to check that the transformation operator Eq.(\ref{C3}) is
Hermitian $\widehat{R}=\widehat{R}^{\dag }$ . Using the transformation
relations \cite{GuAnn}\cite{LLM}

\begin{eqnarray*}
\widehat{R}\widehat{S}_{+}\widehat{R}^{-1} &=&\widehat{S}_{+}\cosh ^{2}\frac{%
\eta }{2}+\widehat{S}_{-}\sinh ^{2}\frac{\eta }{2}-i\widehat{S}_{z}\sinh
\eta , \\
\widehat{R}\widehat{S}_{-}\widehat{R}^{-1} &=&\widehat{S}_{-}\cosh ^{2}\frac{%
\eta }{2}-\widehat{S}_{+}\sinh ^{2}\frac{\eta }{2}-i\widehat{S}_{z}\sinh
\eta , \\
\widehat{R}\widehat{S}_{z}\widehat{R}^{-1} &=&\widehat{S}_{z}\cosh \eta +%
\frac{i}{2}(\widehat{S}_{-}-\widehat{S}_{+})\sinh \eta ,
\end{eqnarray*}

the diagonalized Hamiltonian is indeed Hermitian%
\begin{equation}
\widetilde{H}=\widehat{R}\widehat{H}\widehat{R}^{-1}=\widehat{R}^{-1}%
\widehat{H}^{\dag }\widehat{R}=2i\Gamma _{I}\widehat{S}_{z},  \label{C2}
\end{equation}%
with an effective imaginary frequency
\begin{equation}
\Gamma _{I}=\sqrt{\Omega ^{2}-G^{2}}.  \label{C4}
\end{equation}%
The non-Hermitian coupling constant $G$ decreases the slope of potential
well and the effective frequency. There exists an exceptional point%
\begin{equation*}
G_{c}=\Omega ,
\end{equation*}%
where all eigenstates are degenerate with vanishing eigenvalue. The pending
parameter $\eta $ is determined from the equation%
\begin{equation}
\sinh \eta =\pm \frac{G}{\sqrt{\Omega ^{2}-G^{2}}}  \label{D2}
\end{equation}%
which for a real value of $\eta $ is valid below the exceptional point $%
G<G_{c}$. The Hamiltonian Eq.(\ref{C2}) describes the particle in a deformed
inverted potential well with the reduced slope, which tends to zero at the
exceptional point. The variation of potential-well shape is illustrated in
Fig.1.

The eigenstates of transformed Hamiltonian are obviously%
\begin{equation*}
\widetilde{H}|\widetilde{u}_{n}\rangle _{\gamma }=\pm i\Gamma _{I}\left( n+%
\frac{1}{2}\right) |\widetilde{u}_{n}\rangle _{\gamma },
\end{equation*}%
respectively for the "ket" and "bra" states, $\gamma =r,l.$ In our case it
is simply the eigenstate of imaginary-frequency boson-number $\widehat{n}%
_{I} $
\begin{equation*}
|\widetilde{u}_{n}\rangle _{\gamma }=|n\rangle _{\gamma },
\end{equation*}%
defined in equation Eq.(\ref{B6}). The eigenstates of the original
non-Hermitian Hamiltonians in equations Eq.(\ref{F},\ref{F3}) are obtained
respectively by the inverse transforms
\begin{equation*}
|u_{n}\rangle _{r}=\widehat{R}^{-1}|n\rangle _{r},\quad |u_{n}\rangle _{l}=%
\widehat{R}|n\rangle _{l},
\end{equation*}%
with corresponding eigenvalues
\begin{equation*}
\varepsilon _{n}=i\Gamma _{I}\left( n+\frac{1}{2}\right) ,
\end{equation*}%
and%
\begin{equation*}
\varepsilon _{n}^{\ast }=-i\Gamma _{I}\left( n+\frac{1}{2}\right) .
\end{equation*}

\begin{figure}[tbp]
\includegraphics[width=3.5in]{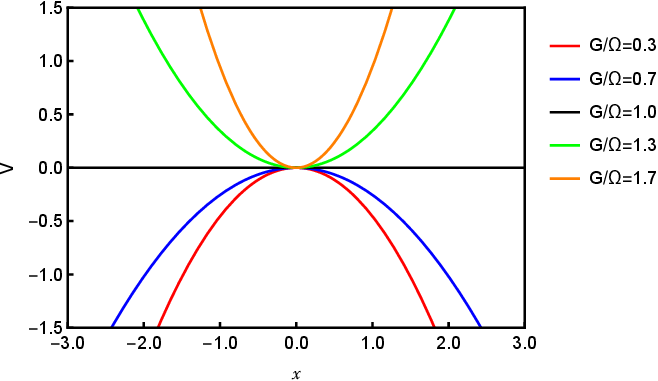}
\caption{The interaction-constant dependence of inverted potential-well
shape for $G=0.3$ (red curve), $0.7$ $\left( \text{blue}\right) $ in the
unit $\Omega $. The potential vanishes at the exceptional point $G_{c}=1.0$ $%
\left( \text{black}\right) $ and becomes normal well beyond $G_{c}$ for $%
G=1.3$ $\left( \text{green}\right) $, $1.7$ $\left( \text{orange}\right) $.}
\end{figure}
The lowest-layer energy spectrum with respect to interaction constant $G$ is
displayed in Fig.2 for $n=0$ (black)$,1$ (blue)$,2$ (red). All eigenvalues
vanish at the exceptional point $G_{c}$, which separates the real and
imaginary domains of energy spectrum. We emphasize that the open orbits
below the exceptional point $G_{c}$ are also quantized, however with two
branches of imaginary discrete-eigenvalues.

\begin{figure}[tbp]
\includegraphics[width=3in]{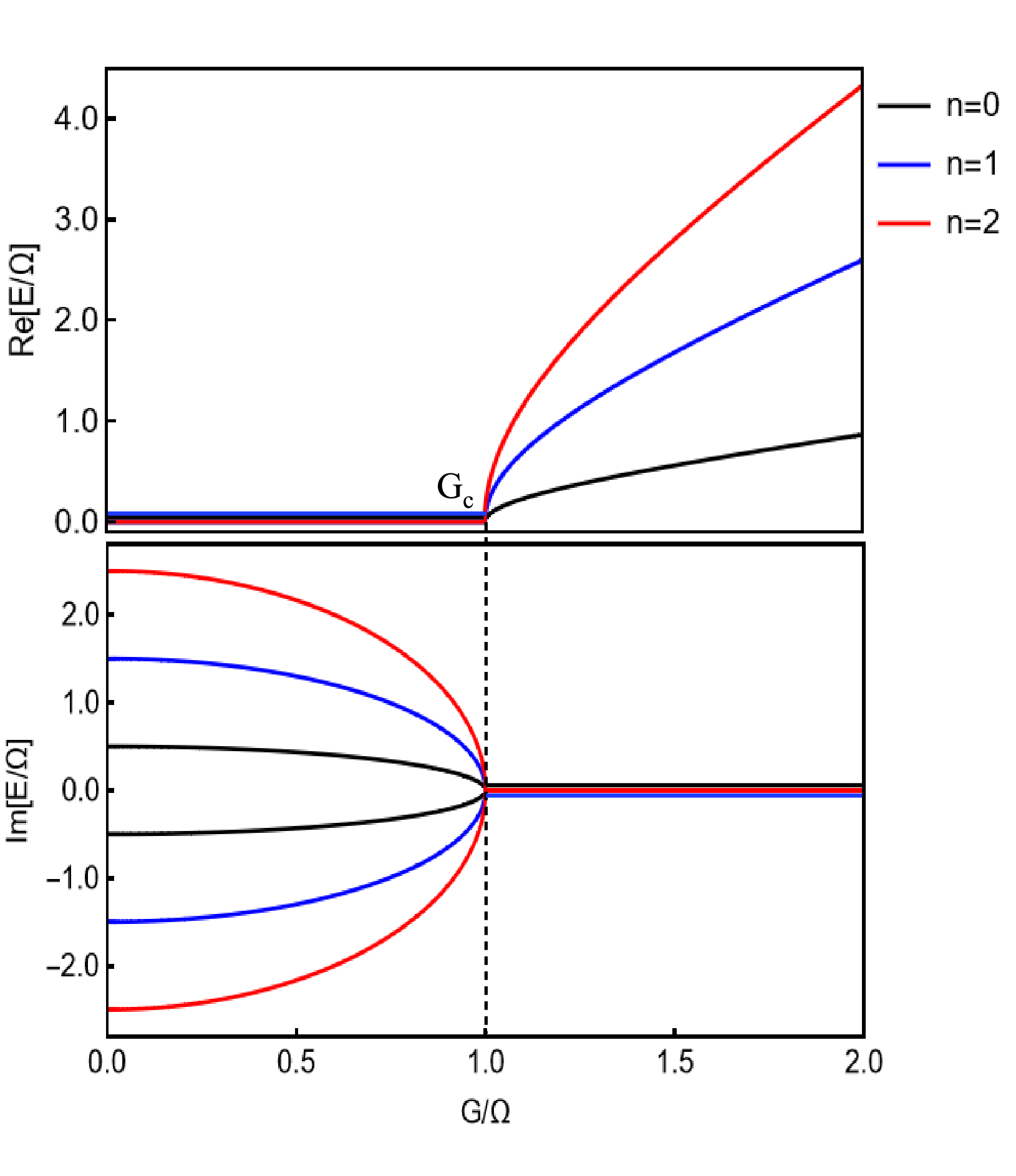}
\caption{The lowest layers ($n=0,1,2)$ of energy spectrum as functions of
the interaction constant $G$. The imaginary eigenvalues below $G_{c}$ have
two branches respectively for the "ket" and "bra" states and vanish at $%
G_{c} $. The spectrum becomes real beyond the exceptional point.}
\end{figure}

With the coordinate representation of $SU(1,1)$ generators $\widehat{S}_{+}$
and $\widehat{S}_{-}$ in equations Eqs.(\ref{G},\ref{C}) the eigenfunctions
of "bra" and "ket" states are obtained as

\begin{eqnarray*}
u_{r,n}\left( x\right) &=&e^{-\frac{\eta }{2}\left( x^{2}-\frac{d^{2}}{dx^{2}%
}\right) }\psi _{r,n}\left( x,\Gamma _{I}\right) , \\
u_{l,n}\left( x\right) &=&e^{\frac{\eta }{2}\left( x^{2}-\frac{d^{2}}{dx^{2}}%
\right) }\psi _{l,n}\left( x,\Gamma _{I}\right) ,
\end{eqnarray*}%
in which the value of parameter $\eta $ is solved from Eq.(\ref{D2}). The
eigenfunctions $\psi _{\gamma ,n}\left( x,\Gamma _{I}\right) =\langle x|%
\widetilde{u}_{n}\rangle _{\gamma }$ for $\gamma =r,l$ \ are the same as in
the equation Eq.(\ref{C1}), however with the reduced frequency $\Gamma _{I}$
instead of $\Omega $. The state density probability is%
\begin{eqnarray*}
\rho _{r,l} &=&\langle x|u_{r,n}\rangle \langle u_{l,n}|x\rangle =\psi
_{r,n}\left( x,\Gamma _{I}\right) \psi _{l,n}^{\ast }\left( x,\Gamma
_{I}\right) , \\
\rho _{l,r} &=&\rho _{r,l}^{\ast }.
\end{eqnarray*}

Beyond $G_{c}$ the potential becomes a normal oscillator-well as shown in
Fig.(1). By analytical continuation extension the transformed Hamiltonian
describes a normal oscillator%
\begin{equation}
\widetilde{H}_{ext}=2\Gamma \widehat{S}_{z},  \label{EX}
\end{equation}%
with%
\begin{equation*}
\Gamma =\sqrt{G^{2}-\Omega ^{2}}.
\end{equation*}%
The $SU(1,1)$ generators in this region are realized by the normal boson
operators $\widehat{a}$, $\widehat{a}^{\dag }$
\begin{equation*}
\widehat{S}_{z}=\frac{1}{2}\left( \widehat{n}+\frac{1}{2}\right) ,\quad
\widehat{n}=\widehat{a}^{\dag }\widehat{a}
\end{equation*}%
where
\begin{equation*}
\widehat{a}=\frac{1}{\sqrt{2}}\left( \widehat{x}+i\widehat{p}\right) ,%
\widehat{a}^{\dag }=\frac{1}{\sqrt{2}}\left( \widehat{x}-i\widehat{p}\right)
.
\end{equation*}%
The commutation relation Eq.(\ref{E}) of $SU(1,1)$ generators is invariant
under the realization of normal boson-operators.

The eigenstates of transformed Hamiltonian Eq.(\ref{EX}) are that of normal
oscillator%
\begin{equation*}
\widetilde{H}_{ext}|n\rangle =\Gamma \left( n+\frac{1}{2}\right) |n\rangle ,
\end{equation*}%
where $\widehat{n}|n\rangle =n|n\rangle $.

\section{Quantum-classical correspondence for the Hamiltonian of inverted
potential well with Non-Hermitian interaction}

The classical counterpart of the quantum Hamiltonian Eq.(\ref{F}) in the
realization of imaginary-frequency boson operators Eq.(\ref{C}),Eq.(\ref{G})
is seen to be%
\begin{equation}
H\left( x,p\right) =\Omega \left( \frac{p^{2}}{2}-\frac{x^{2}}{2}\right)
-iGxp,  \label{C9}
\end{equation}
which is a complex function of canonical variables $x$, $p$ corresponding to
the non-Hermitian interaction. We are going to find what is the classical
counterpart of the transformed Hermitian Hamiltonian Eq.(\ref{C2}). To this
end we adopt a canonical transformation of variables%
\begin{eqnarray}
x &=&X\cosh \frac{\eta }{2}-iP\sinh \frac{\eta }{2},  \notag \\
p &=&iX\sinh \frac{\eta }{2}+P\cosh \frac{\eta }{2},  \label{C6}
\end{eqnarray}%
in which parameter $\eta $ is to be determined. Substituting the canonical
variable transformation Eq.(\ref{C6}) into the phase space Lagrangian
\begin{equation}
\mathcal{L=}\overset{.}{x}p-H\left( x,p\right)  \label{C7}
\end{equation}%
with the classical Hamiltonian given by Eq.(\ref{C9}) yields
\begin{equation}
\mathcal{L}\left( X,P\right) =\mathcal{L}^{\prime }\left( X,P\right) +\frac{%
dF(X,P)}{dt},  \label{C8}
\end{equation}%
where $F(X,P)$ is called the gauge or generating function. Two Lagrangians $%
\mathcal{L}\left( X,P\right) $ and
\begin{equation*}
\mathcal{L}^{\prime }\left( X,P\right) =\overset{.}{X}P-H^{\prime }\left(
X,P\right)
\end{equation*}%
differing by a total time derivative of canonical-variable function $F(X,P)$
are gauge equivalent \cite{LW2023}, since they give rise to the same
equation of motion. The gauge transformation of the Hamiltonian Eq.(\ref{C9}%
) is found as

\begin{equation}
H^{\prime }=\Gamma _{I}\left( \frac{P^{2}}{2}-\frac{X^{2}}{2}\right) ,
\label{C10}
\end{equation}%
under the condition that the pending parameter $\eta $ is determined by the
equation Eq.(\ref{D2}). The Hamiltonian $H^{\prime }$, which indeed is a
real function corresponding to the Hermitian Hamiltonian, describes a
particle in an inverted potential well. It is exactly the classical
counterpart of quantum version $\widetilde{H}$ given in Eq.(\ref{C2}). The
gauge function%
\begin{equation*}
F=\frac{1}{2}\left[ \frac{i}{2}\left( P^{2}-X^{2}\right) -XP\right] \frac{G}{%
\Gamma _{I}}+\frac{1}{2}XP
\end{equation*}%
is however a complex function of the canonical variables. The
quantum-classical correspondence holds strictly \cite{XIN14,GuAnn} for the
Hamiltonian of imaginary spectrum with non-Hermitian interaction.

\section{Conclusion and discussion}

\ The hermiticity of Hamiltonian is not a necessary condition to possess
real spectrum. It is insufficient either. The Hermitian Hamiltonian of a
conservative system can also have complex eigenvalues, if the potential is
unbounded from below. The complex eigenvalues of metastable states were
studied long ago with the semiclassical method \cite{LM92,LM94}. Rigorous
eigenvalues and states are still lacking. The Hermitian Hamiltonian with
inverted potential well is solved in the algebraic method by means of
imaginary-frequency boson operators. The imaginary spectrum is derived
analytically along with a dual-set of eigenstates corresponding respectively
to the complex conjugate eigenvalues. The density-operators are
non-Hermitian invariants, which give rise to the density probability
conservation in agreement with the basic requirement of quantum mechanics.
The $SU(1,1)$ generator $\widehat{S}_{z}$, which possesses always real
spectrum determined by the commutation relation, however is non-Hermitian in
the realization of imaginary-frequency boson operators. An interesting
observation is that the Hermitian Hamiltonian possesses imaginary
eigenvalues while the spectrum of non-Hermitian operator $\widehat{S}_{z}$
is real. It is well known that the spectrum of the pseudo-Hermitian
Hamiltonians is not necessarily real in the whole region of parameter value.
It approaches a complex domain at the exceptional point when the coupling
constant of non-Hermitian interaction increases. Different from the common
belief we find the opposite direction of transition from imaginary to real
domains of eigenvalues at the exceptional point. In the considered $SU(1,1)$
Hamiltonian the non-Hermitian interaction decreases the slope of inverted
potential well. The potential-well slope vanishes at the exceptional point,
where all eigenstates are degenerate with zero eigenvalues. Beyond the
exceptional point the potential becomes normal well with a real spectrum.
Although the $SU(1,1)$ generator $\widehat{S}_{z}$ possesses real spectrum
always, the boson-operator realization of it can be either Hermitian with
the normal boson or non-Hermitian with the imaginary-frequency boson
operators. The quantum-classical one to one correspondence exists strictly
for the Hermitian and non-Hermitian Hamiltonians.

\bigskip

\section{Appendix}

\bigskip

\emph{Hamiltonian of inverted potential:}

The Hamiltonian for a particle of unit mass in an inverted potential well
reads%
\begin{equation}
\widehat{H}=\frac{\widehat{p}^{2}}{2}-\frac{1}{2}\Omega ^{2}\widehat{x}^{2}
\label{A2}
\end{equation}%
with the conventional unit $\hbar =1$. In the dimensionless variables with $%
\widehat{p}$ replaced by $\widehat{p}/\sqrt{\Omega }$and $\widehat{x}$ by $%
\sqrt{\Omega }\widehat{x}$ we have the Hamiltonian Eq.(\ref{A}). The
Hamiltonian Eq.(\ref{A2}) can be regarded as a harmonic oscillator of
imaginary frequency,%
\begin{equation*}
\widehat{H}=\frac{\widehat{p}^{2}}{2}+\frac{1}{2}\left( i\Omega \right) ^{2}%
\widehat{x}^{2}.
\end{equation*}

\bigskip

\emph{Imaginary frequency polynomials:}

Using the commutation relation of boson operators Eq.(\ref{H}) it is easy to
verify
\begin{equation*}
\widehat{n}_{I}\widehat{b}_{-}|n\rangle _{r}=\left( n-1\right) \widehat{b}%
_{-}|n\rangle _{r}.
\end{equation*}%
Thus $\widehat{b}_{-}$ is a lowering operator for the "ket" state. We assume
that
\begin{equation}
\widehat{b}_{-}|n\rangle _{r}=c_{n-}^{r}|n-1\rangle _{r},  \label{R}
\end{equation}%
in which the coefficient $c_{n-}^{r}$ is to be determined. $\widehat{b}_{+}$
acts as a rising operator, since
\begin{equation*}
\widehat{n}_{I}\widehat{b}_{+}|n-1\rangle _{r}=n\widehat{b}_{+}|n-1\rangle
_{r}.
\end{equation*}%
We may define%
\begin{equation*}
\widehat{b}_{+}|n-1\rangle _{r}=c_{\left( n-1\right) +}^{r}|n\rangle _{r}.
\end{equation*}%
Acting the raising operator on the Eq.(\ref{R}) yields
\begin{equation*}
\widehat{n}_{I}|n\rangle _{r}=c_{n-}^{r}c_{\left( n-1\right) +}^{r}|n\rangle
_{r},
\end{equation*}%
and then we have from the orthogonal condition Eq.(\ref{B3}) the recurrence
relation
\begin{equation*}
c_{n-}^{r}c_{\left( n-1\right) +}^{r}=n.
\end{equation*}%
Repeating the same procedure on the states $|n-1\rangle _{r}$, $|n-2\rangle
_{r}$ ; $|n-2\rangle _{r}$, $|n-3\rangle _{r}$ ;$\cdot \cdot \cdot ;$we
obtain

\begin{equation*}
c_{\left( n-1\right) -}^{r}c_{\left( n-2\right) +}^{r}=n-1;
\end{equation*}%
\begin{equation*}
c_{\left( n-2\right) -}^{r}c_{\left( n-3\right) +}^{r}=n-2;
\end{equation*}%
\begin{equation*}
\cdot \cdot \cdot \cdot \cdot \cdot
\end{equation*}%
\begin{equation*}
c_{1-}^{r}c_{0+}^{r}=1.
\end{equation*}%
For the boson number ground-state we require%
\begin{equation}
\widehat{b}_{-}|0\rangle _{r}=0,  \label{R2}
\end{equation}%
and%
\begin{equation*}
c_{0-}^{r}=0.
\end{equation*}%
The coefficients are determined as%
\begin{equation*}
c_{n-}^{r}=c_{\left( n-1\right) +}^{r}=\sqrt{n};c_{\left( n-1\right)
-}^{r}=c_{\left( n-2\right) +}^{r}=\sqrt{n-1};\cdot \cdot \cdot
;c_{1-}^{r}=c_{0+}^{r}=1.
\end{equation*}%
Thus the $n$th "ket" state can be generated from ground state with the
rising operator%
\begin{equation*}
|n\rangle _{r}=\frac{(\widehat{b}_{+})^{n}}{c_{\left( n-1\right)
+}^{r}...c_{0+}^{r}}|0\rangle _{r}=\frac{(\widehat{b}_{+})^{n}}{\sqrt{n!}}%
|0\rangle _{r}.
\end{equation*}%
For the "bra" states situation is opposite that $\widehat{b}_{-}$ acts as a
raising operator
\begin{equation*}
\widehat{n}_{I}^{\dag }\widehat{b}_{-}|n\rangle _{l}=\left( n+1\right)
\widehat{b}_{-}|n\rangle _{l},
\end{equation*}%
and thus%
\begin{equation*}
\widehat{b}_{-}|n\rangle _{l}=c_{n-}^{l}|n+1\rangle _{l}.
\end{equation*}%
While $\widehat{b}_{+}$ is lowering operator
\begin{equation*}
\widehat{n}_{I}^{\dag }\widehat{b}_{+}|n\rangle _{l}=\left( n-1\right)
\widehat{b}_{+}|n\rangle _{l},
\end{equation*}%
and%
\begin{equation}
\widehat{b}_{+}|n\rangle _{l}=c_{n+}^{l}|n-1\rangle _{l}.  \label{R1}
\end{equation}%
The recurrence relation for the "bra" states is obtained from the equations
Eq.(\ref{B2},\ref{R1}) as
\begin{equation*}
n|n\rangle _{l}=\widehat{n}^{\dag }|n\rangle _{l}=-c_{n+}^{l}c_{\left(
n-1\right) -}^{l}|n\rangle _{l},
\end{equation*}%
and from the orthogonal condition Eq.(\ref{B3}) we have
\begin{equation*}
c_{n+}^{l}c_{\left( n-1\right) -}^{l}=-n.
\end{equation*}%
Repeating the same procedure on the states $|n-1\rangle _{l}$, $|n-2\rangle
_{l}$ ; $|n-2\rangle _{l}$, $|n-3\rangle _{l}$ ;$\cdot \cdot \cdot ;$the
result is
\begin{equation*}
c_{(n-1)+}^{l}c_{\left( n-2\right) -}^{l}=-\left( n-1\right) ;
\end{equation*}%
\begin{equation*}
\cdot \cdot \cdot \cdot \cdot \cdot
\end{equation*}%
\begin{equation*}
c_{1+}^{l}c_{0-}^{l}=-1.
\end{equation*}%
For the ground state it is%
\begin{equation}
\widehat{b}_{+}|0\rangle _{l}=0,  \label{R3}
\end{equation}%
and%
\begin{equation*}
c_{0+}^{l}=0.
\end{equation*}%
The solutions of coefficients are seen to be%
\begin{eqnarray*}
c_{n+}^{l} &=&c_{\left( n-1\right) -}^{l}=-i\sqrt{n},c_{(n-1)+}^{l}=c_{%
\left( n-2\right) -}^{l}=-i\sqrt{n-1}, \\
\cdot \cdot \cdot ,c_{1+}^{l} &=&c_{0-}^{l}=-i,
\end{eqnarray*}%
from which the $n$th "bra" state can be generated from the ground state such
as%
\begin{equation*}
|n\rangle _{l}=\frac{\left( \widehat{b}_{-}\right) ^{n}}{(-i)^{n}\sqrt{n!}}%
|0\rangle _{l}.
\end{equation*}

In coordinate representation the ground-state equation Eq.(\ref{R2}) is
\begin{equation*}
\left( x-i\frac{d}{dx}\right) \psi _{r,0}=0,
\end{equation*}%
from which the ground "ket" state is found as%
\begin{equation*}
\psi _{r,0}=\frac{1}{\sqrt{N_{r}}}e^{-i\frac{x^{2}}{2}},
\end{equation*}%
where $N$ is the normalization constant to be determined from the
orthonormal condition Eq.(\ref{B3}) between "bra" and "ket" states. The
arbitrary order eigenfunctions can be generated from the ground state such
as
\begin{equation*}
\psi _{r,n}=\frac{1}{\sqrt{N_{r}}\sqrt{n!}}\left( \sqrt{\frac{i}{2}}\right)
^{n}\left( x-i\frac{d}{dx}\right) ^{n}e^{-i\frac{x^{2}}{2}}.
\end{equation*}

With the same procedure we have the "bra" ground-state equation Eq.(\ref{R3}%
)
\begin{equation*}
\left( x+i\frac{d}{dx}\right) \psi _{l,0}=0,
\end{equation*}%
in the coordinate representation and the wave function%
\begin{equation*}
\psi _{l,0}=\frac{1}{\sqrt{N_{l}}}e^{i\frac{x^{2}}{2}}.
\end{equation*}%
The arbitrary-order wave function of "bra" state is%
\begin{equation*}
\psi _{l,n}=\frac{1}{(-i)^{n}\sqrt{n!}\sqrt{N_{l}}}\left( \sqrt{\frac{i}{2}}%
\right) ^{n}\left( x+i\frac{d}{dx}\right) ^{n}e^{i\frac{x^{2}}{2}}.
\end{equation*}%
The normalization constant can be determined from the orthonormal condition
Eq.(\ref{B3}) between "bra" and "ket" states
\begin{equation*}
\int \psi _{l,0}^{\ast }\psi _{r,0}dx=\frac{1}{\sqrt{N_{l}^{\ast }N_{r}}}%
\int e^{-ix^{2}}dx=1.
\end{equation*}%
The normalization constant
\begin{eqnarray*}
N_{r} &=&\sqrt{\frac{\pi }{i}}, \\
N_{l} &=&\sqrt{\frac{\pi }{-i}}
\end{eqnarray*}%
is then obtained by means of the imaginary integral measure as in the path
integral of quantum mechanics (see for example \cite{FH65}\cite{LW2023}).
The normalized wave functions are given by equation Eq.(\ref{C1}) in Sec.II.


\bigskip

\bigskip

\begin{thebibliography}{99}
\bibitem{CM98} C. M. Bender, S. Boettcher, Phys. Rev. Lett. \textbf{80},
5243 (1998).

\bibitem{CM99} C. M. Bender, S. Boettcher, and P. Meisinger, J. Math. Phys.
\textbf{40}, 2201 (1999).

\bibitem{CM02} C. M. Bender, D. C. Brody, and H. F. Jones, Phys. Rev. Lett.
\textbf{89}, 270401 (2002).

\bibitem{GuResult} Y. Gu, X. M. Bai, X. L. Hao, and J.-Q. Liang, Results.
Phys. \textbf{38}, 105561 (2022).

\bibitem{GuAnn} Y. Gu, X. L. Hao, and J.-Q. Liang, Ann. Phys.(Berlin).
\textbf{534}, 2200069 (2022).

\bibitem{LiuResult} N. Liu, S. Huang, and J.-Q. Liang, Results. Phys.
\textbf{40}, 105813 (2022).

\bibitem{Mosta1} A. Mostafazadeh, Nuclear. Physics. B. \textbf{640}, 419
(2002).

\bibitem{Mosta2} A. Mostafazadeh, J. Math. Phys. \textbf{43}, 205 (2002).

\bibitem{Mosta3} A. Mostafazadeh, Phys.\ Lett. B. \textbf{650}, 208 (2007).

\bibitem{LiuPysica} N. Liu, Y. Gu, and J.-Q. Liang, Phys. Scr. \textbf{98},
035109 (2023).

\bibitem{Amaa} N. Amaaouche, M. Sekhri, R. Zerimeche, M. Mustaphaa, and
J.-Q. Liang, Physics. open. \textbf{13}, 100126 (2022).

\bibitem{CM04} C. M. Bender, D. C. Brody, and H. F. Jones, Phys. Rev. D.
\textbf{70}, 025001 (2004).

\bibitem{CM07} C. M. Bender, Rep. Prog. Phys. \textbf{70}, 947 (2007).

\bibitem{CM18} C. M. Bender, PT Symmetry: in Quantum and Classical Physics
(World Scientific, Singapore, 2018).

\bibitem{DN18} D. N. Christodoulides, J. Yang, Parity-time symmetry and its
applications (Springer Tracts in Modern Physics, 2018).

\bibitem{CH13} C. Hang, G. Huang, and V. V. Konotop, Phys. Rev. Lett.
\textbf{110}, 083604 (2013).

\bibitem{ZZ16} Z. Y. Zhang, Y. Q. Zhang, J. T. Sheng, L. Yang, M. A. Miri,
D. N. Christodoulides, B. He, Y. P. Zhang, and M. Xiao, Phys. Rev. Lett.
\textbf{117}, 123601 (2016).

\bibitem{SB12} S. Bittner, B. Dietz, U. G\"{u}nther, H. L. Harney, M.
Miski-Oglu, A. Richter, and F. Sch\"{a}fer, Phys. Rev. Lett. \textbf{108},
024101 (2012).

\bibitem{CM13} C. M. Bender, B. K. Berntson, D. Parker, and E. Samuel, Am.
J. Phys. \textbf{81}, 173 (2013).

\bibitem{CE10} C. E. R\"{u}ter, K. G. Makris, R. El-Ganainy, D. N.
Christodoulides, M. Segev, and D. Kip, Nat. Phys. \textbf{6}, 192 (2010).

\bibitem{YW19} Y. Wu, W. Q. Liu, J. P. Geng, X. R. Song, X. Y. Ye, C. K.
Duan, X. Rong, and J. F. Du, Science. \textbf{364}, 878 (2019).

\bibitem{LC14} L. Chang, X. Jiang, S. Hua, C. Yang, J. Wen, L. Jiang, G. Li,
G. Wang, and M. Xiao, Nat. Photon. \textbf{8}, 524 (2014).

\bibitem{PG15} P. G. Kevrekidis, J. Cuevas--Maraver, A. Saxena, F. Cooper,
and A. Khare, Phys. Rev. E. \textbf{92}, 042901 (2015).

\bibitem{HX16} H. Xu, D. Mason, L. Jiang, and J. G. E. Harris, Nature.
\textbf{537}, 80 (2016).

\bibitem{WL17} W. L. Li, C. Li, and H. S. Song, Phys. Rev. A. \textbf{95},
023827 (2017).

\bibitem{SK19} S. K. \"{O}zdemir, S. Rotter, F. Nori, and L. Yang, Nat.
Mater. \textbf{18}, 783 (2019).

\bibitem{JZ18} J. Zhang, B. Peng, S. K. \"{O}zdemir, K. Pichler, D. O.
Krimer, G. Zhao, F. Nori, Y. X. Liu, S. Rotter, and L. Yang, Nat. Photon.
\textbf{12}, 479 (2018).

\bibitem{YZ17} Y. L. Liu, R. Wu, J. Zhang, S. K. \"{O}zdemir, L. Yang, F.
Nori, and Y. X. Liu, Phys. Rev. A. \textbf{95}, 013843 (2017).

\bibitem{FQ18} F. Quijandr\'{l}, U. Naether, S. K. \"{O}zdemir, F. Nori, and
D. Zueco, Phys. Rev. A. \textbf{97}, 053846 (2018).

\bibitem{Liang} J.-Q. Liang, H. J. W. M\"{u}ller-Kirsten, Y. B. Zhang, A. V.
Shurgaia, S. P. Kou, and D. K. Park, Phys. Rev. D. \textbf{62}, 025017
(1999).

\bibitem{LLM} Y. Z. Lai, J.-Q. Liang, H. J. W. M\"{u}ller-Kirsten, and J. G.
Zhou, Phys. Rev. A. \textbf{53}, 3691 (1996).

\bibitem{XIN14} J. L. Xin, J. -Q. Liang, Sci. China. Phys. Mech. Astron.
\textbf{57}, 1504 (2014).

\bibitem{LW2023} J. -Q. Liang and L. F. Wei, New Advances in Quantum Physics
(Third edition) (Science Press, Beijing, 2023).

\bibitem{LM92} J. -Q. Liang and H. J. W. M\"{u}ller-Kirsten, Phys. Rev. D.
\textbf{45}, 2963 (1992)

\bibitem{LM94} J. -Q. Liang and H. J. W. M\"{u}ller-Kirsten, Phys. Rev. D.
\textbf{50}, 6519 (1994)

\bibitem{FH65} R. P. Feynman and A. R. Hibbs, Quantum mechanics and path
integrals, New York: McGraw-Hill, 1955
\end{thebibliography}
\end{document}